\documentclass[aps,prl,twocolumn,showpacs,superscriptaddress,groupedaddress]{revtex4}

\usepackage{amsmath}
\usepackage{latexsym}
\usepackage{float}
\usepackage{amssymb}
\usepackage{graphicx}
\usepackage{epsfig}
\usepackage{color}

\begin{document}

\title{Living clusters and crystals from low density suspensions of active colloids }

\author{B.\ M.\ Mognetti$^\dagger$}
\affiliation{Deparment of Chemistry, University of Cambridge, Lensfield Road, Cambridge, CB2 1EW, United Kingdom}
%\affiliation{$^{*}$These authors contributed equally}

\author{A.\ \v{S}ari\'c$^\dagger$}
\affiliation{Department of Chemistry, Columbia University, 3000 Broadway, New York, NY 10027}
\thanks{These authors contributed equally}
%\affiliation{$^{*}$These authors contributed equally}

\author{S.\ Angioletti-Uberti}
\affiliation{Deparment of Chemistry, University of Cambridge, Lensfield Road, Cambridge, CB2 1EW, United Kingdom}
\affiliation{Institute of Physics, Humboldt Universit\"{a}t zu Berlin, Newtonstr. 15, 12489 Berlin, Germany}

\author{A.\ Cacciuto}
\affiliation{Department of Chemistry, Columbia University, 3000 Broadway, New York, NY 10027}

\author{ C.\ Valeriani}
\email{cvaleriani@quim.ucm.es}
\thanks{Corresponding author}

\affiliation{Departamento de Quimica Fisica I, Facultad de Ciencias Quımicas, Universidad Complutense de Madrid, 28040 Madrid, Spain}

\author{D.\ Frenkel} 
\affiliation{Deparment of Chemistry, University of Cambridge, Lensfield Road, Cambridge, CB2 1EW, United Kingdom}

\begin{abstract}

Recent studies aimed at investigating artificial analogues of bacterial colonies have shown that low--density suspensions of self--propelled 
particles confined in two dimensions can assemble into finite aggregates that merge and split, but have a typical size that remains constant (living clusters). 
In this Letter we address the problem of the formation of living clusters and crystals of active particles in three dimensions. 
We study two systems: self-propelled particles interacting via a generic attractive potential and colloids that can move towards each other as a result of active agents (e.g.\ by molecular 
motors). In both cases fluid--like `living' clusters form. We explain this general feature in terms of the balance between active forces and regression to thermodynamic equilibrium. This balance can be quantified in terms of a { dimensionless} number that allows us to collapse the observed clustering behaviour onto a universal curve. We also discuss how active motion affects the kinetics of crystal formation.

\end{abstract}

\maketitle

% PACS numbers: 05.70.Ce, 64.70.F-, 64.75.Cd, 02.70.Tt
%Electronic mail: bm411@cam.ac.uk, as3499@columbia.edu,  cvaleriani@quim.ucm.es, df246@cam.ac.uk

Active systems consume energy that keeps them in an out--of--equilibrium state \cite{Zia,ActiveGeneral}. 
This is typical for many biologically relevant systems which, by exploiting chemical energy, can self-organise into complex structures that lack any equilibrium counterpart. 
Examples are abundant and exist at different length-scales: from cytoskeleton remodulation during cell mitosis \cite{GeneralBio} to swarming phenomena in micro-swimmers or flocks of birds \cite{Animal,Robot}. 
The similarity of the patterns displayed by these systems lead many to address the general principles behind their formation using simple models \cite{Vicsek}. %such as for example the Vicksek model \cite{Vicsek}. 
 The reproducibility and robustness of the phenomena under a variety of external conditions motivated a large body of research on self-assembly 
of active particle as a possible strategy towards the fabrication of new functional nano- and mesoscopic structures  \cite{3dChip}. In this respect there have been several efforts to study active self-organisation in a tightly controlled environment; the most studied systems being self--propelled particles \cite{Vicsek,SPGeneral,add1,add2,add3} and active gels (e.g.\ \cite{Cytoskeleton}).  

Recently, two experimental groups have shown how two--dimensional suspensions of self--propelled  { (SP)} colloids 
(moving through the consumption of an appropriate `fuel') self--assemble into dynamic clusters that constantly join and split, recombining with each other to reach a steady-state \cite{Lyon2012,NY2013}.
A general { understanding} of active cluster formation is lacking, but { different explanations} have been proposed.
The authors of Ref.\ \cite{Lyon2012} argued that a net attraction between colloids is responsible for clustering. This attraction is due to non-uniformity in the chemical fuel concentration between two colloids. 
This was confirmed by Ref.\ \cite{NY2013}, that also found a $1/d^2$ dependence of the particle-particle attraction ( $d$ being the inter-particle distance).
The formation of finite-size clusters has been also observed in low density bacteria/polymer suspensions, at intermediate polymer concentrations just before phase separation \cite{add2}.
{ However, in other recent studies (e.g.\ \cite{add1,add3,Syracuse2012,Lowen}) clusters were shown to form due to dynamic instabilities, 
hence attraction cannot be singled out as the only cause promoting aggregation in active systems.}

Doubts remain over the nature of the driving forces behind clustering observed in different experiments. Furthermore, it is not clear whether this phenomenon is specific to these systems or a generic feature
of non-equilibrium.

To address these questions, in this Letter we report { and explain} the formation of living clusters in two 
very different active systems  and { at arbitrarily low packing fraction}. 
The first system consists of SP particles similar to those used in Refs.\ \cite{Lyon2012,NY2013}  with an added isotropic attraction. 
%\cite{Lyon2012,NY2013} 
%(Fig.\ \ref{Fig1}$a$).
The second  are hard spheres that, when closer than a given range, are brought together by molecular motors  acting as dynamical cross links between `tracks' (e.g. microtubules) grafted to the particles. The model was inspired by recent work in which centrosomes~\cite{SyntheticCentrosome} were self--assembled with the microtubule polarity constrained to point inward/outward 
\footnote{Molecular motors are not present in the system of Ref.\ \cite{SyntheticCentrosome}.  Here we use the concept of synthetic centrosomes as a possible realisation of SD particles.}. 
Note that in this model, motors' action leads to a net attraction between particles. We refer to these particles as self-displacing (SD). 
To highlight the difference between the models, consider that
 in the absence of activity and at the packing fractions used in this study (two orders 
of magnitude below hard sphere freezing) SP particles would condense, while SD particles 
would remain in the gas phase. Furthermore, the  role of activity in the two systems is inverted; 
in the SP model it tends to break up clusters, whereas for SD particles it drives their formation, and splitting is instead due to diffusion.

{ By comparing these two different systems, we show} that the assembly of finite sized clusters at low packing fraction is a generic feature of suspensions of active colloids 
and it is not limited to the systems studied in Refs~ \cite{Lyon2012,NY2013}. Moreover, we demonstrate  
that the formation of finite aggregates can be interpreted as a competition between equilibrium and active forces. 
%Finally, we propose a simple self--tuning argument that explains the robustness of the cluster formation observed in \cite{Lyon2012,NY2013}.

\begin{figure}[h]
\vspace{0.5cm}
\includegraphics[angle=0,scale=0.3,clip=]{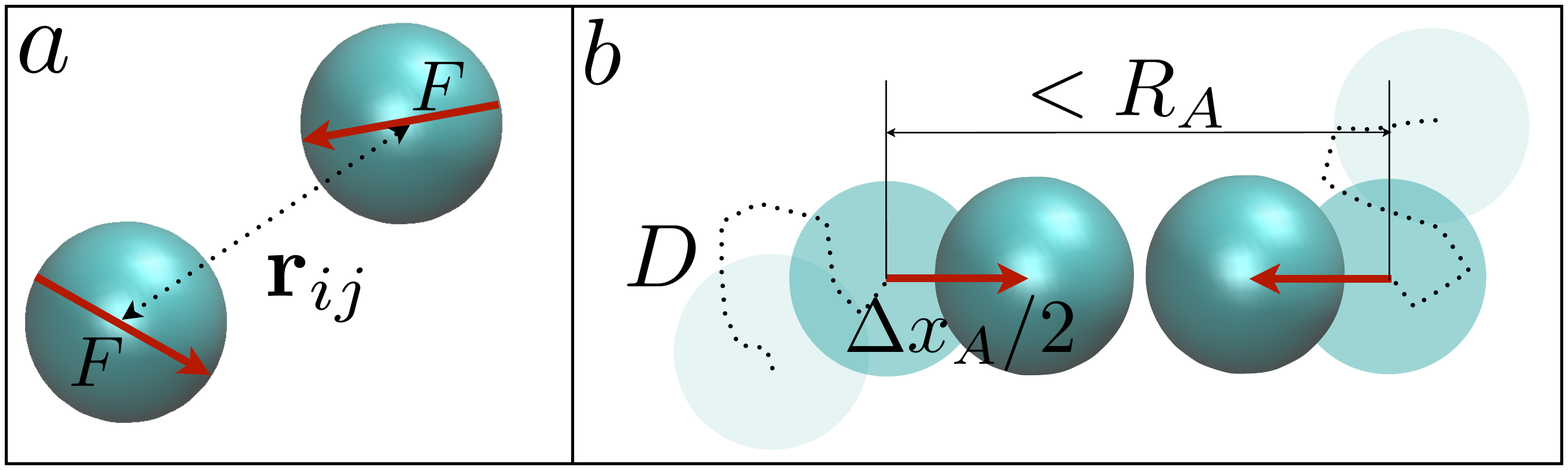} 
\vspace{-3.5cm}
\caption{$a)$ Self--propelled (SP) particles (with propulsion force $F$) interacting via a Lennard--Jones potential. 
$b)$ Self-displacing (SD) particles moving towards each other over a distance $\Delta x_A$ with a rate $\nu$ if they are within a distance $R_A$.  Particles also undergo diffusion ($D$).
Depending on the values of $\Delta x_A$, $\nu$ and $D$, either clustering or collisions  dominates. }\label{Fig1} 
%\vspace{-0.1cm}
\end{figure}

{\em Simulation details}. 
Self-propelled colloids are modelled as spherical particles  of radius $\sigma_p=10\sigma$ (where $\sigma$ is the MD unit of length) interacting via a Lennard-Jones potential
\begin{eqnarray}
 V(r_{ij})=4\epsilon \left[ \left( {\sigma_p \over r_{ij} } \right)^{12}-\left( {\sigma_p \over r_{ij} } \right)^6 \right],
\end{eqnarray}
truncated and shifted at 25$\sigma_p$. Self--propulsion is implemented by adding a constant force $\boldsymbol{F}$ acting along a predefined  axis through the particle (see Fig.\ \ref{Fig1}$a$).
To enable the rotation of the axis of the colloid, and therefore the direction of its propulsion, two small ideal particles with diameter $\sigma$ are placed inside each colloid along its axis and positioned 4.5$\sigma$ symmetrically with respect to the center of the particle, forming a rigid body with the colloid. Though the small particles do not interact with any other particle in the system, they are however subjected to the thermal fluctuations induced by the bath, thus leading to a net Brownian rotation of their axis.
The motion of each particle is governed by the Langevin equation: 
%m \ddot\boldsymbol{r_i}= -\sum\limits_{i \neq j}\partial V_{ij} \over \partial \boldsymbol{r_{ij}} -\zeta \dot{\boldsymbol{r_i}} + \boldsymbol{F_i} + \boldsymbol{F_{R,i}} \\

\begin{equation}
m \ddot{\boldsymbol{r_i}}= - \sum\limits_{j \neq i} {\partial V(r_{ij}) \over \partial \boldsymbol{r_i}} -\zeta \dot{\boldsymbol{r_i}} + \boldsymbol{F_i} + \boldsymbol{F_{R,i}} \\
\label{eq:langevin}
\end{equation}
where $m$ is the sphere's mass, set to 1, and $\zeta$ is the friction coefficient ($\zeta=m\gamma$ with damping coefficient $\gamma$). $F_{R,i}=\sqrt{k_{\rm B}T\zeta}R_i(t)$ is the the random force due to the solvent, where $R_i(t)$ is a stationary Gaussian noise with zero mean and variance $\left\langle R_i(t)R_j(t')\right\rangle =\delta(t-t')\delta_{ij}$. The  damping parameter is $\gamma=\tau^{-1}$, $\tau$ being the MD time unit. 
Simulations were performed at two different packing fractions $\phi=0.01$ and $\phi=0.1$, using the code LAMMPS \cite{LAMMPS} with a total number of particles of $N_\mathrm{part} = 1728$  for at least $15 \cdot 10^6$ time-steps (being   $\Delta t =0.008\, \tau$).
Each simulation was repeated between 4 and 16 times with different initial random velocities. 

SD colloids are modelled as hard spheres with diameter $\sigma$ following Brownian motion with diffusion constant $D$. Whenever two (randomly chosen) particles are closer than  a given distance ($R_A$), they can be self--displaced toward each other along the direction joining their centres.
The activity of SD colloids is then specified by the range $R_A$ within which particles can be cross--linked, the rate $\nu$ at which pulling events occur and $\Delta x_A$, the size of the active displacement (see Fig.\ \ref{Fig1}$b$). In this work we set $R_A=2\sigma$. The  dynamics is implemented using a Monte Carlo algorithm generating diffusive or attractive displacements with the correct frequencies.
For a given `pulling' rate $\nu$  { we randomly select a colloid $i$ and one of its neighbours}  $j$. Particle $i$ and $j$ are then moved toward each other over a distance $min[\Delta x_A, r_{ij} -\sigma]$, unless this motion results in an overlap with a third colloid. 
The active displacement does not alter the centre of mass of the two colloids.
%\marginpar{How about small $\nu$?}
{ If $\Delta x_A \geq R_A-\sigma$ the colloids are displaced to their closest possible configuration. We call this the "processive limit" because molecular motors would drag the two colloids until the end of the microtubule is reached. Simulations with SD colloids were performed with $N_{part}=1000$ and $N_{part}=2000$.}

{\em Aggregation of living clusters}. 
One of the most interesting results of our simulations is that, despite being fundamentally different,   both models lead to the formation of disordered `living' aggregates.  { The degree of clustering is measured by the }
\begin{equation}
\Theta = 1 - { 1 \over \langle S_\mathrm{clust} \rangle } \, ,
\label{eq:degree-clustering}
\end{equation}
 where $\langle S_\mathrm{clust} \rangle$ is the average number of particles in a cluster. 
$\Theta$ ranges from 0 in the gas phase ($\langle S_\mathrm{clust}\rangle$=1) to $1-1/N_\mathrm{part}\approx 1$ when all atoms belong to a single cluster ($\langle S_\mathrm{clust}\rangle=N_\mathrm{part}$).
Two particles are defined as clustered whenever $r_{ij} < 1.2\cdot\sigma_p$ (for SP particles)  or  if $r_{ij}<R_A$ (for SD particles). 
The number and morphology of the clusters depend on the relative ratio between equilibrium and out-of-equilibrium control parameters, as measured by a dimensionless quantity 
which we call the propensity for aggregation $P_\mathrm{agg}$. In the case of SP colloids $P_\mathrm{agg}$ is simply the ratio between the strength of the 
attraction between particles and the propelling force:  $P_\mathrm{agg,SP} \equiv \varepsilon/(F\sigma_p)$, while for SD colloids, $P_\mathrm{agg}$ is the ratio between the 
times to move colloids actively and diffusively over a distance of $R_A-\sigma$: $P_\mathrm{agg,SD}=\nu/[2D/(R_A-\sigma)^2]$.

\begin{figure}[h!]
\vspace{0.1cm}
\includegraphics[angle=0,scale=0.3,clip=]{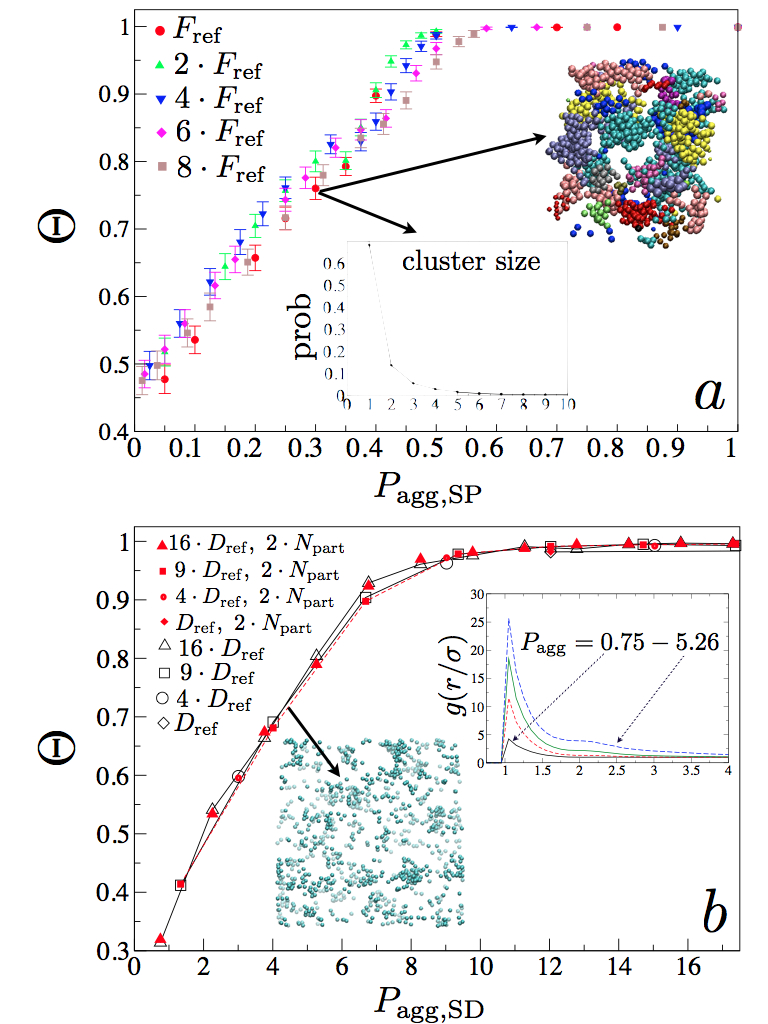} 
%\vspace{0.5cm}
\caption{ { Degree of clustering $\Theta$ \textit{versus} aggregation propensity} ({see text for the definitions}) for the SP ($a$) and the SD systems ($b$). { Results for several propulsion forces $F$ ($a$) and diffusion coefficients $D$ ($b$) collapse onto two universal curves}. 
{The insets represent typical snapshots of living clusters, the cluster size distribution function ($a$), and the pair distribution function ($b$) at significant $P_\mathrm{agg}$ where finite clusters form}. Particles belonging to different clusters in (a) are colored differently and, for clarity, monomers and small clusters are not shown.
%and the probability distribution of the cluster size at a  given 
%$\epsilon F/\sigma$ and $\nu/ [2 D/ (R_A-\sigma)^2)]$, respectively. 
In the SP case $\phi=0.1$ and $F_\mathrm{ref} \sigma_p=10k_{\rm B}T$, while in the SD case $d_\mathrm{cm}/R_A=2$ ($\phi=0.008$), $N_\mathrm{part}=1000$ and $D_\mathrm{ref}=0.01\sigma/\tau_{MC}$, $\tau_{MC}$ being the simulation time unit.}\label{Fig2} 
\vspace{0.cm}
\end{figure}

 Fig.\ \ref{Fig2}$a$ shows  the number of clusters per particles in the SP colloidal suspension for different values of $\epsilon$ and $F$. These data represent the steady-state value for the cluster distribution, 
 to which different initial states converge after a short transient time. Strikingly, all data collapse onto a single master curve when plotted as a function of $P_\mathrm{agg,SP}$. 
{ Furthermore, the figure shows a linear increase in the degree of clustering with decreasing activity (i.e. increasing $P_{agg,SP}$)},
 %Furthermore, the figure shows a linear decrease in the number of clusters with the strength of the activity,
which supports the simple picture of the mechanism of cluster formation in terms of two competing terms:
the LJ attraction drives aggregation, whilst activity  counteracts the formation of large clusters.

%{\em What is the Peclet number of self--propelled colloids?} We now try to rationalise our SP
 %results in view of experimental results. A key difference between our system and Refs.\ %\cite{Lyon2012,NY2013} is that for chemically driven SP particles both the propulsion %($F_\mathrm{exp}$) and the colloidal attraction ($\varepsilon_\mathrm{exp}$) depend on the
%fuel concentration $c$. The latter is due to the phoretic motion \cite{phoretic}
%Defining $c_{12}(\bf{x})$ the density profile of the chemical fuel constrained by the presence of %two SP particles at fixed position (${\bf x}_1$ and ${\bf x}_2$), the osmotic interaction between %colloids can be written as $\int \mathrm{d} {\bf x} (\vec \nabla c_\mathrm{12} ({\bf x}))^2$. 
%Notice we are assuming that the diffusivity of the fuel is big and that the fuel concentration %satisfies the Dirichlet problem $\Delta c_{12}=0$ with $c_{12}=0$ at the the colloid surface and %$c_{12}(\infty)=c$. 
%Since $c_{12}(\bf{x}) =c \phi({\bf x};{\bf x}_1,{\bf x}_2) $ (where $\phi$ does not depend on %$c$), and given that for an isolated colloids $F\sim c$, it follows that the net interaction %between two SP particles scale with $\epsilon_\mathrm{exp} \sim F_\mathrm{exp}^2$, and as a %consequence the size of a cluster, at small activity, increases linearly with $F_\mathrm{exp}$ %consistently with the results reported in Ref.\ \cite{Lyon2012}. 
%This confirms the important role played by attraction in the mechanism driving the cluster %formation and explains a self--tuning mechanism between attraction and propulsion arising in %experiments. 

Fig.\ \ref{Fig2}$b$ shows the same analysis for SD colloids 
in the processive limit ($\Delta x_A>R_A-\sigma$) at a colloidal packing 
fraction of $8\cdot 10^{-3}$. { In this case clustering is observed when activity increases. The trend is opposite to that observed in Fig.\ \ref{Fig2}$a$ for SP particles because, as discussed above, in the SD suspension activity is responsible for aggregation, whereas for SP particles activity is the limiting factor for cluster formation.
} Increasing the colloidal diffusivity increases the probability of a colloid to detach from the cluster. Interestingly, as in the SP system, the number of clusters exhibits a linear dependence on $P_\mathrm{agg,SD}$. 
However, as we approach the limit in which a single big cluster forms (see Fig.\ \ref{Fig2}), we
observe that the relaxation time required to reach the steady state increases sensibly, 
suggesting the presence of a phase transition. We defer this issue to a future  study.

In both systems we find that the cluster size distribution appears to follow a power law 
(see  insets in Fig.\ \ref{Fig2}$a$). Clusters coexist  with a monomers--rich  gas phase.
The clusters grow and shrink dynamically. An analysis of  the pair correlation function (inset of 
Fig.\ \ref{Fig2}$b$) does not reveal any sign of structural order at longer range than the 
typical cluster size. { Notice that living clusters form in the region where $P_\mathrm{agg,SP}\approx 0.1-0.5$ and $P_\mathrm{agg,SD}\approx 1-5$, i.e. where active and thermal forces  have comparable magnitude.
This highlights the synergic nature of these non--equilibrium aggregates. 
{This power-law behavior for cluster-size distributions has also been observed as a function of energy-intake and particle density  in a recent study on active Brownian particles~\cite{lobaskin}.}

Interestingly, when SP particles were confined to move in 2D, { for moderate $P_\mathrm{agg,SP}$} we obtained living clusters with crystalline order analogous to those observed in the experiments of Ref.\ \cite{NY2013}. {However, we never observed living clusters with crystalline order in 3D in the regime 
where active and thermal forces have comparable magnitude.}

{\em Driven phase diagram of the SD suspension}
In the SD suspension, activity is controlled not only by the  pulling rate but also by the size of the active displacement $\Delta x_A$. 
In Fig.\ \ref{Fig3}$a$ we show how clustering is altered by the latter parameter. 
{ As expected, decreasing $\Delta x_A$  the degree of clustering can be kept constant by increasing $\nu$. }
\begin{figure}[h!]
\vspace{-0cm}
\includegraphics[angle=0,scale=0.33]{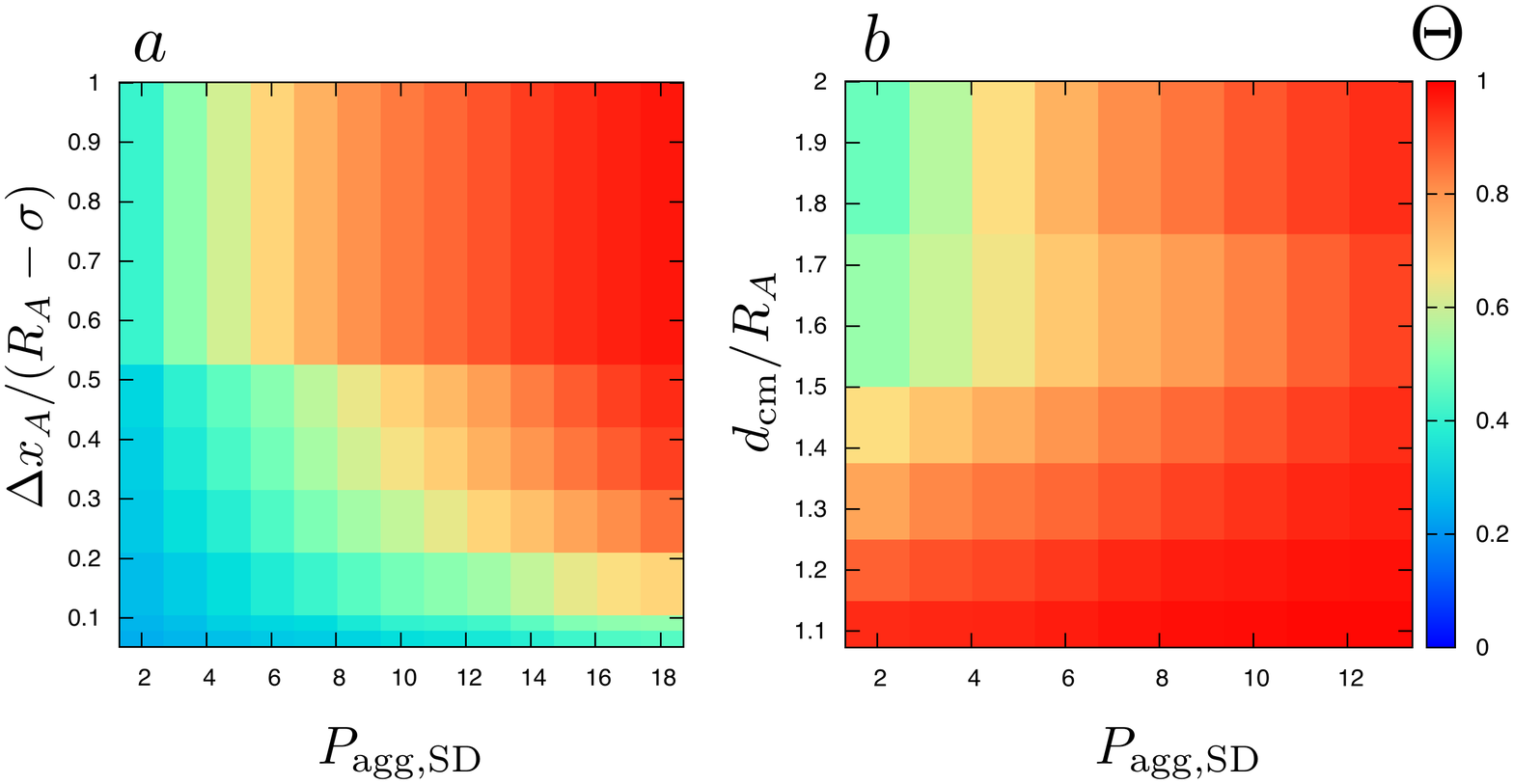} 
\vspace{-2.5cm}
\caption{($a$)  Fraction of clusters in the SD suspension
 as a function of the processivity parameter $\Delta x_A/(R_A-\sigma)$ versus 
 the aggregation propensity $P_\mathrm{agg,SD}$ and  ($b$) as a function of their average distance $d_\mathrm{cm}=\rho^{-1/3}$
 versus $P_\mathrm{agg,SD}$. {In ($a$) $d_\mathrm{cm}/R_A=2$ ($\phi=0.008$).}}
\label{Fig3}
\vspace{0.1cm}
\end{figure}

In Fig.\ \ref{Fig3}$b$ we investigate the change in the number of clusters as a function of  the  number density $\rho$. More specifically we plot it as a function of the average distance between colloids in the bulk, $d_\mathrm{cm} = \rho^{-1/3}$.
Not surprisingly, a larger degree of activity is required to maintain clustering at large values of $d_\mathrm{cm}$. Both Figs.\ \ref{Fig2} and \ref{Fig3} suggest the presence of a dynamical transition in which the number of clusters varies continuously from one to a finite fraction. 
 
{\em Active crystallisation}
Beyond a well-defined onset value of $P_\mathrm{agg}$ we observe cluster crystallisation.
Crystallinity of the aggregates is detected  by evaluating the $q_6$ bond--order parameter and using the criteria described in Ref~\cite{cryst1,cryst2} 
%As $q_6$  can be quite inaccurate for small clusters (due to their large surface-to-volume ratio) we computed the bond order parameter only for  clusters formed at least by $10^2$ particles {\bf BM check for the SD particles.}).  [UNNECESSARY]
Fig.\ \ref{Fig4} shows the fraction of solid-like particles as a function of $P_\mathrm{agg,SP}$ for suspensions of SP ($a$) and SD ($b$) particles. 
\begin{figure}[h!]
\vspace{-0.cm}
\includegraphics[angle=0,scale=0.3,clip=]{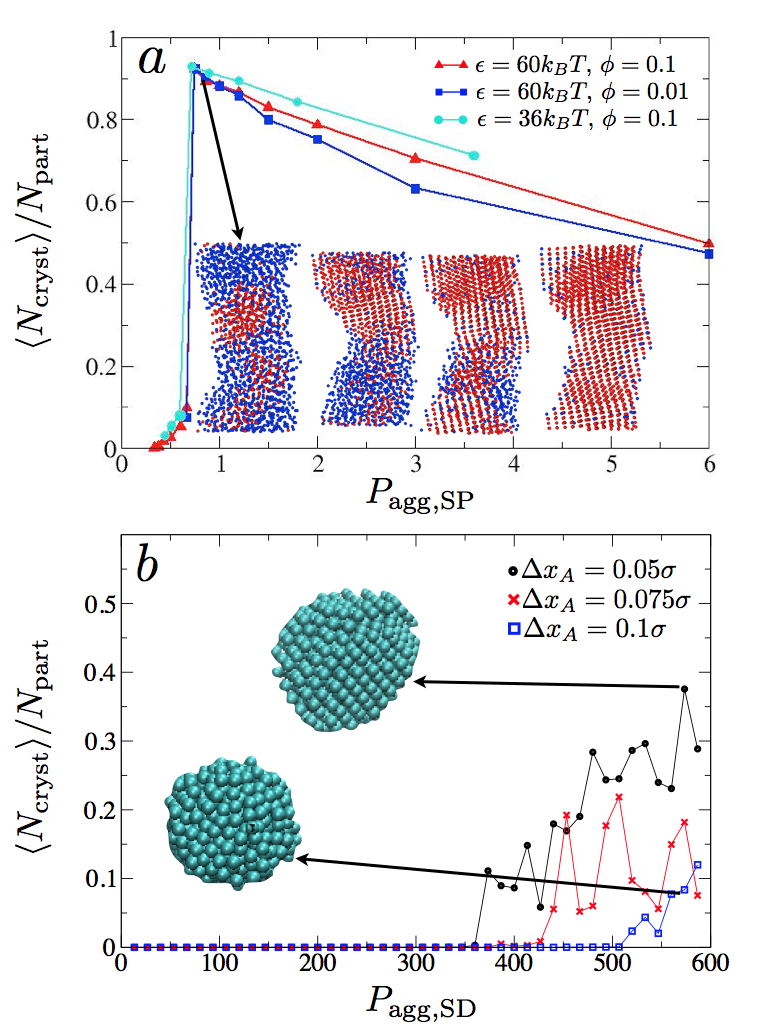} 
\vspace{-0cm}
\caption{Crystallisation of active particles. {Fraction} of solid-like particles in the system versus the { aggregation propensity} in the SP ($a$) and SD ($b$) case. The snapshots in $a)$ show the crystallization process of a single cluster ($\phi=0.1$, $\epsilon=60k_{\rm B}T$ and $F\sigma_p=80k_{\rm B}T$) during the course of a simulation. Fluid-like particles are depicted in blue, solid-like particles in red.  { Crystals span the whole simulation box } 
$b)$ SD particles at low $\Delta x_A/\sigma$ form small faceted crystals that are replaced by arrested aggregated when $\Delta x_A$ increases (inset snapshots). In this case  $d_\mathrm{cm}/R_A=2$ ($\phi=0.008$).}\label{Fig4}
\vspace{-0.cm}
\end{figure}
For SP particles (Fig.\ \ref{Fig4}$a$) the fraction of crystalline particles exhibits a 
non--monotonic behaviour. 
For $\epsilon >> F\sigma_p$ (in our simulations, $\epsilon \geq 30k_{\rm B}T$ and $F\sigma_p \sim 10k_{\rm B}T$), we observe the formation of disordered structures that survive throughout the  simulation. 
Formation of the equilibrium crystalline phase at high values of $\epsilon$ is kinetically hindered by the strong inter-particle attraction, which leads to low diffusivity and hence long equilibration times.
In the opposite regime, when $\epsilon\approx 0.1 F\sigma_p $ { (i.e. $P_{agg,SP} \approx 0.1$ )}, the living fluid aggregates described  in Fig.\ \ref{Fig2} make their appearance. 
At intermediate values of $\epsilon/F\sigma_p$ (see Fig.\ \ref{Fig4}$a$) instead, the balance between particle attraction and activity leads to the formation of crystalline clusters. 
A possible interpretation of these results is that in this intermediate regime, the role of activity is that of speeding-up the kinetics of crystal formation by increasing the diffusivity, allowing a faster annealing of defects. 

It is instructive to relate the simulation parameters in the SP case to the experimental systems discussed in Ref. \cite{Lyon2012,NY2013,Lowen,Syracuse2012}. If a colloid of $\sigma_p=1\mu m$ is propelling with $v \sim 5\mu m/s$, its propulsion force in water is $F \sim 0.05 pN$. Our analysis suggests that  the interaction strength between the colloids  required to form a single  macroscopic fluid cluster should be  $\epsilon \sim 6 k_{\rm B}T$ ( $P_\mathrm{agg,SP} \sim 0.5$). For lower attractions the system forms many living clusters, as typically observed in experiments. According to Fig.\ \ref{Fig4}$a$, a well-ordered crystal of active particles should be expected {within a few seconds} for $\epsilon \sim 12k_{\rm B}T$.

Fig.\ \ref{Fig4}$b$ shows the crystallisation route followed by SD particles. 
In this system, crystals only form for small displacements. When the processivity parameter $\Delta x_A>0.1$, aggregates instead undergo structural arrest before they can rearrange into an ordered structure. 
We believe the reason behind the formation of regular structures with SD particles is similar to the mechanism leading to crystallisation in an external field (e.g.\ \cite{ext1,ext2}). 
However, here the effective external forces push the particles toward the center of a cluster. 
It should be stressed that in both our systems crystallisation typically leads to irreversible clustering rather than to an equilibrium size distribution of living clusters.

{\em Conclusion}. In this Letter we have investigated the formation of living fluid-like clusters and crystals from low density suspensions of active colloids. 
We have studied two different systems: self--propelled particles interacting via an attractive potential and hard spheres in which colloids are  pairwise self--displaced toward each others. In both cases  we observed the formation of living fluid clusters (as in Ref.\cite{Lyon2012,NY2013}) and irreversible crystals of active particles. 
We showed how the formation of finite aggregates can be ascribed to a balanced competition between equilibrium forces and activity, regardless of the nature of the latter . 
Furthermore, we demonstrated that activity aids annealing defects in crystalline clusters. 

{\bf Acknowledgments:} This work was supported by the ERC Advanced Grant 227758, the National Science Foundation under Career Grant No. DMR-0846426, the Wolfson Merit Award 2007/R3 of the Royal Society of London and the EPSRC Programme Grant EP/I001352/1. BMM acknowledge T.\ Curk and A.\ Ballard for useful discussions. C. V. acknowledges financial support from a Juan de la Cierva Fellowship, from the Marie Curie Integration Grant PCIG-GA-2011-303941 ANISOKINEQ, and from the National Project FIS2010-16159. S. A-U acknowledges support from the Alexander von Humboldt Foundation.


\begin{thebibliography}{99}
\bibitem{cryst2}
Two particles $i$ and $j$ are considered bonded if their center of the mass distance is $r_{ij} < 1.25\sigma_p$ (in the case of SP particles) or if $r_{ij} < R_A$ (for SD particles). Therefore, a particle is identified as solidlike if it is bonded to more than 7 neighbors and if $q_6(i)q_6(j) > 0.5$.
\end{thebibliography}
\end{document}